%% file: main.tex
\input{header.tex}

\newcommand{\MG}[1]{\textcolor{black}{#1}}

\definecolor{pant24PeachFuzz}{RGB}{255,190,152}

\IEEEoverridecommandlockouts
\begin{document}
\input{acronyms.tex}

\title{%Random Multiple Access for IoT Services in \\ Coexisting GEO-LEO Satellite Systems Alternative title: 
To Share, or Not to Share: A Study on GEO-LEO Systems for IoT Services with Random Access}

\author{Marcel~Grec, Federico~Clazzer, Israel~Leyva-Mayorga, Andrea~Munari, Gianluigi~Liva, and Petar~Popovski
\thanks{M.~Grec, F.~Clazzer, A.~Munari and G.~Liva are with the Inst.\ of Communications and Navigation, German Aerospace Center (DLR), Wessling, Germany (e-mail: \{marcel.grec, federico.clazzer,  andrea.munari, gianluigi.liva\}@dlr.de).}%
\thanks{I.~Leyva-Mayorga and P.~Popovski are with the Dep. of Electronic Systems, Aalborg University, Aalborg, Denmark. (email: \{ilm, petarp\}@es.aau.dk). This work was partly supported by the Villum Investigator grant ``WATER'' from the Villum Foundation, Denmark.}%
%\thanks{M.~Grec, F.~Clazzer, A.~Munari and G.~Liva acknowledge the financial support by the Federal Ministry of Education and Research of Germany in the programme of ``Souver{\"a}n. Digital. Vernetzt.'' Joint project 6G-RIC, project identification number: 16KISK022.}%
}

\maketitle
\thispagestyle{empty}

\begin{abstract}
\MG{
The increasing number of satellite deployments, both in the low and geostationary Earth orbit
exacerbates the already ongoing scarcity of wireless resources when targeting ubiquitous
connectivity. For the aim of supporting a massive number
of IoT devices characterized by bursty traffic and modern variants of random access,
we pose the following question: \emph{Should competing satellite operators share spectrum
resources or is an exclusive allocation preferable?}
This question is addressed by devising a communication model for two operators which
serve overlapping coverage areas with independent IoT services.
Analytical approximations, validated by Monte Carlo simulations, reveal that spectrum sharing can
yield significant throughput gains for both operators under certain conditions tied to the relative
serviced user populations and coding rates in use. These gains are sensitive also
to the system parameters and may not always render the spectral coexistence mutually advantageous.
Our model captures basic trade-offs in uplink spectrum sharing and provides novel actionable insights
for the design and regulation of future 6G non-terrestrial networks.}
\end{abstract}

\input{notation}

\input{introd}

\input{sys_model}

\input{analytic_approx}

\input{results}

\section{Conclusions}

This paper has studied the problem of whether it is better for two distinct satellite operators -- one employing exclusively \ac{LEO} satellites, while the other, just \ac{GEO} satellites -- to share their originally allocated frequency bands and have their users coexist or to keep their allocated band exclusively for themselves. The use-case under consideration is servicing \ac{IoT} devices in a remote monitoring scenario with bursty traffic which deploys a modern random access policy i.e., \ac{CRDSA}. We have found that for carefully selected channel code rates, each individual \ac{IoT} service \MG{has} a sizeable benefit in the peak throughput by sharing its band. We observed a gain of up to 36.5\% for the \ac{LEO} operator and of up to 60.6\% for the \ac{GEO} operator in peak throughput with respect to the case where each keeps its traffic on the original allocated band.

\vspace{-0.1cm}

\bibliographystyle{IEEEtran}
\bibliography{IEEEabrv, refs}

\end{document}

%% file: header.tex
\documentclass[conference, 10pt, twocolumn]{IEEEtran}
\usepackage{cite}
\usepackage{amsmath,amssymb,amsfonts}
\usepackage{graphicx}
\usepackage{xcolor}
\usepackage[nolist]{acronym}
\usepackage{comment}
\usepackage{mathrsfs,bm} 
\usepackage{scalerel}
\usepackage[caption=false]{subfig}
\usepackage{tikz}
\usepackage{pgfplots}
\usepackage{tabu}
\usepackage{flushend}
\usepackage{mathtools}
\usepackage{soul}
\usepackage{svg}
\usepackage[acronyms]{glossaries}
\usepackage{multirow}
\usepackage{cases}

%% file: acronyms.tex
%\iffalse
\begin{acronym}
        \acro{2SRA}{two-step random access}
        \acro{4-QAM}{4-quadrature amplitude modulation}
        \acro{16-QAM}{16-quadrature amplitude modulation}

        \acro{4SRA}{four-step random access}
        \acro{5GNR}{5G New Radio}
        \acro{3GPP}{3rd Generation Partnership Project}
        % A
        \acro{ADC}{analog-to-digital converter}
        \acro{AWGN}{additive white Gaussian noise}
        % B
        \acro{BER}{bit error rate}
        \acro{BLER}{block error rate}
        \acro{BP}{belief propagation}
        \acro{BS}{base station}
        % C
        \acro{CCS}{coded compressed sensing}
        \acro{CAZAC}{constant amplitude zero autocorrelation}
        \acro{c.c.u.}{complex channel use}
        \acro{CP}{cyclic prefix}
        \acro{CRDSA}{contention resolution diversity slotted ALOHA}
        \acro{CS}{compressed sensing}
        \acro{CSA}{coded slotted Aloha}
        \acro{c.u.}{channel use}
        % D
        \acro{DAC}{digital-to-analog converter}
        \acro{DD}{delay-Doppler}
        \acro{DE}{density evolution}
        \acro{DFT}{discrete Fourier transform}
        \acro{DZT}{discrete Zak transform}
        % E
        \acro{EP}{embedded pilot}
        % F
        % G
        \acro{GEO}{geostationary-Earth orbit}
        % H
        \acro{HAP}{high-altitude platform}
        % I
        \acro{ICI}{inter-carrier interference}
        \acro{ICSI}{ideal channel state information}
        \acro{IDMA}{interleaver division multiple access}
        \acro{IDFT}{inverse discrete Fourier transform}
        \acro{IDZT}{inverse discrete Zak transform}
        \acro{IFFT}{inverse fast Fourier transform}
        \acro{IoT}{Internet of Things}
        \acro{IRIS$^2$}{Infrastructure for Resilience, Interconnectivity and Security by Satellite}
        \acro{IRSA}{irregular repetition slotted Aloha}
        \acro{ISI}{inter-symbol interference}
        % J
        % K
        % L
	\acro{LDPC}{low-density parity-check}
        \acro{LEO}{low-Earth orbit}
        \acro{LLR}{log-likelihood ratio}
        \acro{LMMSE}{linear minimum mean squared error}
        \acro{LTE}{Long Term Evolution}
        % M
        \acro{MAC}{multiple access}
        \acro{MEO}{medium-Earth orbit}
        \acro{MIMO}{multiple-input multiple-output}
        \acro{MMSE}{minimum mean square error}
        \acro{MPR}{multi-packet reception}
        \acro{MRC}{maximal-ratio combining}
        \acro{MTC}{machine-type communication}
        \acro{MTO}{many-to-one}
        % N
        \acro{NB-IoT}{Narrowband IoT}
        \acro{NMSE}{normalized mean squared error}
        \acro{NR}{new radio}
        \acro{NTN}{non-terrestrial network}
        % O
        \acro{OFDM}{orthogonal frequency-division multiplexing}
        \acro{OMP}{orthogonal matching pursuit}
        \acro{OTFS}{orthogonal time frequency space}
        \acro{OTO}{one-to-one}
        % P
        \acro{p.m.f.}{probability mass function}
        \acro{PAM}{pulse-amplitude modulation}
        \acro{PAPR}{peak-to-average power ratio}
        \acro{PBCH}{physical broadcast channel}
        \acro{PDSCH}{physical downlink shared channel}
        \acro{PO}{PUSCH occasion}
        \acro{PRACH}{physical random access channel}
        \acro{PRB}{physical resource block}
        \acro{PSS}{primary synchronization signal}
        \acro{PUPE}{per-user probability of error}
        \acro{PUSCH}{physical uplink shared channel}
        % Q
        \acro{QPSK}{quadrature phase shift keying}
        \acro{QAM}{quadrature amplitude modulation}

        % R
        \acro{RA}{random access}
        \acro{RCP}{reduced cyclic prefix}
        \acro{RCU}{random coding union}
        % S
        \acro{SB-IDMA}{sparse block interleaver division multiple access}
        \acro{SCL}{successive cancellation list}
        \acro{SCS}{sub-carrier spacing}
        \acro{SE}{spectral efficiency}
        \acro{SIC}{successive interference cancellation}
        \acro{SINR}{signal-to-noise plus interference ratio}
        \acro{SNR}{signal-to-noise ratio}
        \acro{SPARC}{sparse regression code}
        \acro{S1D}{superimposed pilot}
        \acro{SSS}{secondary synchronization signal}
        % T
        \acro{TA}{time advance}
        \acro{TF}{time-frequency}
        \acro{TBS}{transport block size}
        \acro{TIN}{treat-interference-as-noise}
        % U
        \acro{UT}{user terminal}
        \acro{UMAC}{unsourced multiple access}
        % V
        % W
        % X
        % Y
        % Z
        \acro{ZC}{Zadoff-Chu}
        \acro{ZF}{zero forcing}
        \acro{ZP}{zero padding}
\end{acronym}
%\fi

%\newacronym{aloha}{ALOHA}{Additive Links On-line Hawaii Area}
%\newacronym{sa}{SA}{slotted ALOHA}
%\newacronym{irsa}{IRSA}{Irregular Repetition Slotted ALOHA}
%\newacronym{geo}{GEO}{geostationary Earth orbit satellite}
%\newacronym{leo}{LEO}{low Earth orbit satellite}
%\newacronym{iot}{IoT}{Internet of Things}
%\newacronym{redcap}{RedCap}{Reduced Capability}
%\newacronym{6g}{6G}{6th generation of wireless communications}
%\newacronym{snr}{SNR}{signal-to-noise power ratio}
%\newacronym{awgn}{AWGN}{additive white Gaussian noise}
%\newacronym{lna}{LNA}{low-noise amplifier}
%\newacronym{isl}{ISL}{inter-satellite link}
%\newacronym{rtt}{RTT}{round-trip time}
%\newacronym{mac}{MAC}{medium access control}
%\newacronym{qos}{QoS}{Quality of Service}
%\newacronym{noma}{NOMA}{non-orthogonal multiple-access}
%\newacronym{crdsa}{CRDSA}{Contention Resolution Diversity Slotted ALOHA}
%\newacronym{sic}{SIC}{successive interference cancellation}
%\newacronym{plr}{PLR}{packet loss rate}
%\newacronym{iris}{IRIS$^2$}{Infrastructure for Resilience, Interconnectivity and Security by Satellite}
%\newacronym{ofdm}{OFDM}{orthogonal frequency-division multiplexing}

%% file: notation.tex
%%%% Environments
\newtheorem{definition}{Definition}
\newtheorem{prop}{Proposition}

\renewcommand{\Pr}{\ensuremath{\text{Pr}}}

% general and MAC
\newcommand{\saa}{\mathsf{a}}
\newcommand{\sbb}{\mathsf{b}}

\newcommand{\geo}{\mathtt{G}}
\newcommand{\leo}{\mathtt{L}}
\newcommand{\x}{v}
\newcommand{\y}{w}

\newcommand{\ban}{B}
\newcommand{\sysban}{\ban_s}

\newcommand{\nsl}{m}
\newcommand{\nslleo}{{\nsl}_\leo}
\newcommand{\nslgeo}{{\nsl}_\geo}

\newcommand{\us}{u}
\newcommand{\usleo}{\us_\leo}
\newcommand{\usgeo}{\us_\geo}

\newcommand{\usfrac}{\beta}

\newcommand{\lo}{G}
\newcommand{\loth}{\lo^*}

% physical layer and decoding
\newcommand{\numBit}{k}

\newcommand{\rate}{R}
\newcommand{\rategeo}{\rate_\geo}
\newcommand{\rateleo}{\rate_\leo}

\newcommand{\mutInf}{\mathsf{I}}
\newcommand{\avMutInf}{\bar{\mutInf}}

\newcommand{\Pw}{P}
\newcommand{\Ns}{N}

\newcommand{\nc}{h}

% link budget
\newcommand{\AG}{\mathsf{G}}
\newcommand{\FSL}{F}
\newcommand{\Bc}{k_B}
\newcommand{\SNT}{T}

% metrics
\newcommand{\GP}{S}
\newcommand{\GPn}{\bar{\GP}}
\newcommand{\GPmax}{\GP^*}
\newcommand{\GPmaxleoSCA}{\GP^{*}_{\leo,(\saa)}}
\newcommand{\GPmaxgeoSCA}{\GP^{*}_{\geo,(\saa)}}
\newcommand{\GPmaxleoSCB}{\GP^{*}_{\leo,(\sbb)}}
\newcommand{\GPmaxgeoSCB}{\GP^{*}_{\geo,(\sbb)}}
\newcommand{\GPmaxAp}{\hat{\GPmax}}
\newcommand{\GPleo}{\GP_\leo}
\newcommand{\GPgeo}{\GP_\geo}
\newcommand{\ps}{p_s}
\newcommand{\psx}{p_{s,\x}}

% DE
\newcommand{\dg}{d}
\newcommand{\dgM}{\dg_M}
\newcommand{\ud}{\lambda}
\newcommand{\sd}{\rho}
\newcommand{\fb}{\mathsf{f}_\mathsf{u}}
\newcommand{\fs}{\mathsf{f}_\mathsf{s}}

% input and metrics
\newcommand{\lob}{\lo_\sbb}
\newcommand{\loh}{\lo_h}
\newcommand{\lol}{\lo_l}
\newcommand{\loa}{\lo_\saa}
\newcommand{\Olo}{\lo^*}
\newcommand{\Oloh}{\Olo_h}
\newcommand{\Olol}{\Olo_l}

\newcommand{\plr}{\mathsf{p_l}}
\newcommand{\plrb}{\mathsf{p_l^{(\sbb)}}}
\newcommand{\plra}{\mathsf{p_l^{(\saa)}}}
\newcommand{\plrah}{\mathsf{p_{l,h}}}
\newcommand{\plral}{\mathsf{p_{l,l}}}

\newcommand{\tr}{\mathsf{T}}
\newcommand{\tra}{\tr_\saa}
\newcommand{\trb}{\tr_\sbb}
\newcommand{\trah}{\tr_h}
\newcommand{\tral}{\tr_l}
\newcommand{\Otr}{\tr^*}

\newcommand{\se}{\mathsf{S}}
\newcommand{\sea}{\se_{\saa}}
\newcommand{\seb}{\se_{\sbb}}

\newcommand{\rateb}{\rate_{\sbb}}
\newcommand{\ratel}{\rate_{l}}
\newcommand{\rateh}{\rate_{h}}

% system parameters

\newcommand{\banH}{\mathsf{\ban}_h}
\newcommand{\banL}{\mathsf{\ban}_l}
\newcommand{\fth}{\alpha}
\newcommand{\Ofth}{\alpha^*}

\newcommand{\nuh}{\mathsf{\bar u}_h}
\newcommand{\nul}{\mathsf{\bar u}_l}
\newcommand{\fuhl}{\beta}

\newcommand{\nslh}{\mathsf{\nsl}_h}
\newcommand{\nsll}{\mathsf{\nsl}_l}

\newcommand{\nsy}{\mathsf{L}}
\newcommand{\tsym}{T_{s}}
\newcommand{\tsymla}{T_{s,l}^{(\saa)}}
\newcommand{\tsymha}{T_{s,h}^{(\saa)}}
\newcommand{\tsymb}{T_{s}^{(\sbb)}}

\newcommand{\jia}{\mathsf{J}_{\saa}}

\newcommand{\rep}{\mathsf{r}}
\newcommand{\nin}{\mathsf{t}}

% physical layer and decoding

\newcommand{\numSym}{n_s}

\newcommand{\rx}{y}
\newcommand{\rxVec}{\bm{\rx}}
\newcommand{\symVal}{x}
\newcommand{\symVec}{\bm{\symVal}}
\newcommand{\intVal}{z}
\newcommand{\intVec}{\bm{\intVal}}
\newcommand{\noise}{n}
\newcommand{\noiseVec}{\bm{\noise}}
\newcommand{\noiseVar}{\sigma^2}

\newcommand{\Pwha}{\mathsf{P}_{h}^{(\saa)}}
\newcommand{\Pwla}{\mathsf{P}_{l}^{(\saa)}}
\newcommand{\Pwhb}{\mathsf{P}_{h}^{(\sbb)}}
\newcommand{\Pwlb}{\mathsf{P}_{l}^{(\sbb)}}
\newcommand{\intPw}{\mathsf{Z}}

\newcommand{\Nsb}{\mathsf{N}_{\sbb}}
\newcommand{\Nsla}{\mathsf{N}_l^{(\saa)}}
\newcommand{\Nsha}{\mathsf{N}_h^{(\saa)}}

\newcommand{\snt}{\mathcal{T}}

\newcommand{\E}{\mathsf{E}}

\newcommand{\sinr}{\gamma}
\newcommand{\sinrlb}{\gamma_l^{(\sbb)}}
\newcommand{\sinrhb}{\gamma_h^{(\sbb)}}
\newcommand{\sinrla}{\gamma_l^{(\saa)}}
\newcommand{\sinrha}{\gamma_h^{(\saa)}}

\newcommand{\rvDec}{D}

% IRSA stuff

\newcommand{\rarr}{$\Rightarrow~$}
\newcommand{\topnot}[2]{{#1}^{(#2)}}
\newcommand{\xtoy}[2]{{#1}\rightarrow {#2}}

\newcommand{\bw}{B}
\newcommand{\bwfrac}{\beta}

\newcommand{\uspopa}{\mathcal{A}}
\newcommand{\uspopb}{\mathcal{B}}

\newcommand{\modidx}{M}
\newcommand{\infobits}{k}
\newcommand{\rxpow}{\power_r}
\newcommand{\txpow}{\power_t}
\newcommand{\antgain}{G}
\newcommand{\anteff}{\eta}
\newcommand{\leoanteff}{\eta_{\leo}}
\newcommand{\geoanteff}{\eta_{\geo}}
\newcommand{\anttemp}{\mathcal{T}}
\newcommand{\leoanttemp}{\anttemp_{\leo}}
\newcommand{\geoanttemp}{\anttemp_{\geo}}
\newcommand{\rxnfig}{\mathcal{N}_r}
\newcommand{\txantgain}{\antgain_t}
\newcommand{\rxantgainleo}{\antgain_\leo}
\newcommand{\rxantgaingeo}{\antgain_\geo}
\newcommand{\pathlossleo}{L_\leo}
\newcommand{\pathlossgeo}{L_\geo}

\newcommand{\georatemin}{\rate_{\geo, min}}
\newcommand{\leoratemin}{\rate_{\leo, min}}
\newcommand{\georatemax}{\rate_{\geo, max}}
\newcommand{\leoratemax}{\rate_{\leo, max}}
\newcommand{\ratefrac}{\alpha}
\newcommand{\numsgeo}{n_\geo}
\newcommand{\numsleo}{n_\leo}
\newcommand{\noisetempleo}{T_\leo}
\newcommand{\noisetempgeo}{T_\geo}
\newcommand{\noisepow}{N}
\newcommand{\symsinr}{\Omega}
\newcommand{\mutinf}{I}
\newcommand{\aimi}{\bar{\mutinf}}
\newcommand{\symbdur}{T_{s}}
\newcommand{\bolzC}{\bar{k}}

\newcommand{\rxsnr}{\gamma}
\newcommand{\desnr}{\Gamma}

\newcommand{\load}{L}
\newcommand{\loadleo}{\load_\leo}
\newcommand{\loadgeo}{\load_\geo}
\newcommand{\sicrounds}{R_{SIC}}
\newcommand{\numinterf}{h}

\newcommand{\speff}{S}
\newcommand{\thr}{\Theta}
\newcommand{\leoplr}{\plr_{l}}
\newcommand{\geoplr}{\plr_{g}}
\newcommand{\leothr}{\thr_{l}}
\newcommand{\geothr}{\thr_{g}}
\newcommand{\leospeff}{\speff_{l}}
\newcommand{\geospeff}{\speff_{g}}

\newcommand{\carrfreq}{f_c}
\newcommand{\txband}{B}
\newcommand{\txgain}{G_t}

\newcommand{\leominelev}{\epsilon_L}
\newcommand{\geominelev}{\epsilon_G}

\newcommand{\leoalt}{h_L}
\newcommand{\geoalt}{h_G}
\newcommand{\antdiam}{\mathcal{D}}
\newcommand{\leoantdiam}{\antdiam_{\leo}}
\newcommand{\geoantdiam}{\antdiam_{\geo}}

\newcommand{\macframesize}{N_F}
\newcommand{\nsimframes}{M}
\newcommand{\maxsiciter}{\text{\.{N}}}
\newcommand{\minframesize}{\text{\d{M}}}

%% file: introd.tex
\section{Introduction}
\acresetall

Ubiquitous connectivity has \MG{already} been pursued by the \MG{\ac{3GPP}} as part of the 5G standardization activity~\cite{rinaldi2020non}.
\MG{Even so, the effort towards 6G has placed the \ac{NTN} component that entails air and space communications on an integration course into the terrestrial one, resulting in a \emph{3D network}~\cite{guidotti2024role}.}
%The focus has shifted currently towarEven so, the current efforts towards 6G are focusing on integrating the \MG{\ac{NTN}} component -- which entails air and space communications -- into the terrestrial one, resulting in a \emph{3D network}~\cite{guidotti2024role}.
The space segment has recently been characterized by the use of hybrid satellite constellations, often combining \ac{LEO} satellites with either \ac{GEO} or \ac{MEO} satellites (e.g., the merging of Eutelsat with OneWeb or the planned \MG{\ac{IRIS$^2$}} constellation~\cite{IRIS2}). 
Especially in remote natural environments with limited terrestrial coverage, the availability of satellite services becomes essential for providing coverage to \ac{IoT} devices. These devices have typically limited computational capabilities and are required to operate with low energy. Small data fragments are generated sporadically or based on unpredictable events, resulting in a small subset of users active each instant in time. The use of scheduling-based approaches is inefficient as the overhead required to allocate resources for communications to a rapidly varying terminal population can become comparable to the data transmission. Random access protocols \MG{mitigate by design} the need for overhead at the expense of multiple-access interference~\cite{centenaro2021survey}. Also, recent studies show a rapid increase in connected devices~\cite{IoT_forecast22} to enable different \ac{IoT} services concerning logistics, smart farming, health care and automotive.
Hence, orchestrating communication links efficiently for serving heterogeneous \ac{IoT} devices under the constraints of scarce frequency spectrum becomes particularly challenging.
%Hence, new challenges are arising as to how to orchestrate communication links efficiently, that aim at serving heterogeneous \ac{IoT} services especially as the frequency spectrum is becoming increasingly scarce.
Spectrum sharing practices for \ac{IoT} devices have been studied in \cite{zhang2019}. Here, operators try to accommodate \ac{IoT} traffic within the existing licensed band or can resort to using unlicensed spectrum.
The possibility of providing \ac{IoT} connectivity via individual \ac{OFDM} sub-carriers as part of the WiFi broadband signal has been investigated in  \cite{pirayesh2020}.
In \cite{clazzer2024}, a scenario is studied where \ac{IoT} traffic is served via a modern random access scheme by two \ac{NTN} operators, one employing a \ac{LEO} satellite and the other, a high-altitude platform. The paper presents an optimal partitioning of a common frequency band among the two operators in terms of throughput, but also fairness.
Other works dealing with band coexistence or sharing include \cite{popovski2018, qian2021, munari2021}. 

In this paper, we focus on a scenario \MG{
where two competing satellite operators -- one deploying \ac{LEO} satellites, while the other, \ac{GEO} satellites -- provide connectivity to \ac{IoT} devices on ground. These devices activate sporadically, their transmissions are bursty in nature and their access to the medium is done by means of the modern random access scheme \ac{CRDSA}~\cite{Casini2007}.
}
%comprised by two \ac{IoT} services whose connectivity is provided by two competing satellite network operators, the first one deploying \glspl{leo} and, the second one \ac{GEO} satellites. Bursty traffic is generated by a massive number of sporadically active terminals and is served by a modern random access policy, i.e. \ac{CRDSA}~\cite{Casini2007}. %The aim of the 
Our contribution is three-fold. First, we present a simple yet comprehensive system model able to capture the key features of a two-service and two-operator satellite communication network. Secondly, we provide a comprehensive analysis for the scenario where the two satellite networks operate in distinct bands.
\MG{Finally, we present novel conditions in which the satellite operators have mutual benefits in terms of system throughput, when deciding to share their originally allocated frequency bands.}
%Finally, we provide guidelines in which conditions the satellite operators are incentivized to share the bandwidth to increase their system's throughput. 

%% file: sys_model.tex
\section{System Model}
\label{sec:sys_ov}

\begin{figure}
    \centering
    \includegraphics[width=0.8\linewidth]{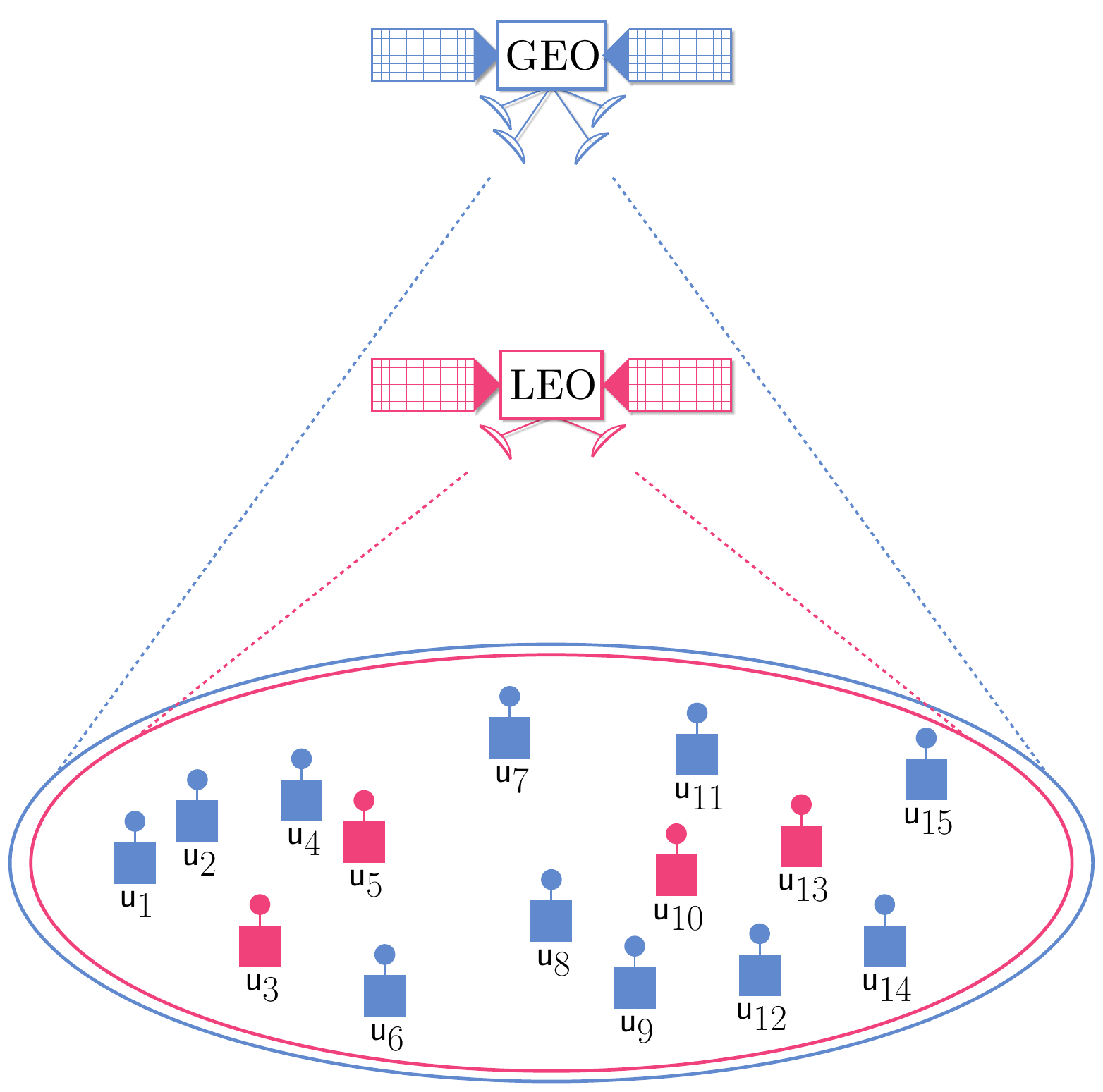}
    \caption{\ac{IoT} users in simultaneous coverage of a \ac{GEO} and a \ac{LEO} satellite.}
    \label{fig:scenario}
\end{figure}

As presented in Fig.~\ref{fig:scenario}, we consider a scenario with two satellite operators serving a common coverage area. The two providers deploy different satellite constellations, one composed only by \ac{LEO} satellites while the other only by \ac{GEO} satellites, and serve two distinct \ac{IoT} services. By assuming a quasi-Earth-fixed cell service link type, and focusing on one specific cell served both by the \ac{LEO} and by one beam of the \ac{GEO}, we consider the setting in which the coverage areas of the two systems coincide. This corresponds to the worst-case scenario with maximal random access competition among the two systems. 

We consider two scenarios. In \emph{scenario ($\saa$)} each of the two satellite operators utilizes a bandwidth $\ban$. In this way, the data transmission of each \ac{IoT} service happens in a different channel, and there is no inter-service multiple-access interference. In \emph{scenario ($\sbb$)} instead, the two operators decide to share both channels so that each system coexists over a common bandwidth $2 \ban$. %In this latter case, we assume that the two channels originally allocated to the two systems can be shared. %Nonetheless, due to limitations in the spectrum allocation, the common bandwidth remains separated in two distinct channels of bandwidth $\sysban/2$ each, due to, for example, discontinuity in the frequency allocation among the two satellite operators.

%-----------------------------------------------
\subsection{Random Access Policy}

\begin{figure}
    \centering
    \includegraphics[width=.9\linewidth]{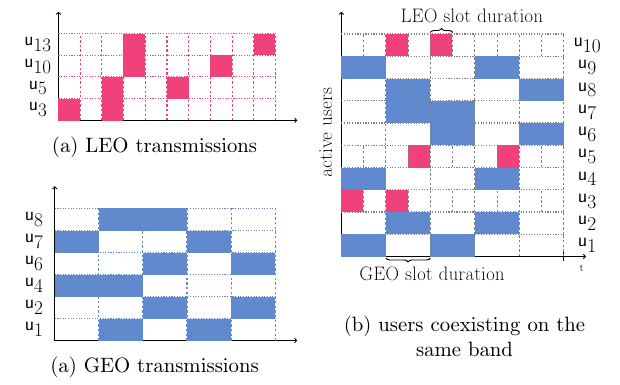}
    %\caption{ (a) corresponds to scenario~($\saa$) and depicts \gls{crdsa} \gls{mac} frames when operators use 2 separate channels, each of bandwidth $\ban$. (b) corresponds to scenario~($\sbb$), treats coexisting operators on each of the 2 available channels. A \gls{mac} frame with mixed traffic is exemplified for one channel of bandwidth~$\ban$.}
    \caption{\MG{\ac{MAC} frames where each operator uses a separate channel of bandwidth $\ban$ are depicted in (a). If operators share the channels, the frames contain mixed traffic on each channel. One such frame is illustrated in (b).}}
    \label{fig:macframes}
\end{figure}

In both scenarios, the chosen \ac{MAC} policy is \ac{CRDSA}. Transmissions are slot-synchronous and users, when active, transmit two copies of their physical layer packets within a \ac{MAC} frame, which is a collection of consecutive time slots. The two packet copies, referred to as replicas, are transmitted by selecting uniformly at random two slots in a \ac{MAC} frame so that no self-interference is caused. \MG{\ac{MAC} frame examples are given in Fig.~\ref{fig:macframes} case (a).} Replicas carry the information about the slot selection enabling the receiver to perform \ac{SIC}. Each active terminal has $\numBit$ information bits to transmit. The replicas are encoded with a Gaussian codebook. To account for the different propagation distances and receiver characteristics, two coding rates (bits/symbol) are \MG{used}: $\rateleo$ for the \ac{LEO} and $\rategeo$ for the \ac{GEO} system. Next, we define $\ratefrac$ as the ratio:
\begin{equation}
    \label{eq:alphadef}
    \ratefrac \triangleq \frac{\rateleo}{\rategeo},
\end{equation}
with $\alpha \in \mathcal{S} \subset \mathbb{N}$ and $\alpha\geq1$. Note that packets are transmitted over the same bandwidth $\ban$ regardless of the scenario and the system considered. Hence, the symbol duration remains fixed and the packet duration for the \ac{GEO} system will be $\ratefrac$~times the duration of the replicas in the \ac{LEO} system. Every replica fits in one \ac{LEO} or \ac{GEO} slot and the \ac{MAC} frame duration is kept the same, \MG{thus} only some integer values of $\alpha$ are allowed.\footnote{In general $\alpha$ can take any positive real value. The condition $\alpha \in \mathcal{S} \subset \mathbb{N}$ is considered to ease the simulations and will be relaxed in future extensions. Note that, due to the different link conditions, depending on the choice of $\rateleo$, $\alpha>1$. This ensures that the packets intended for the \ac{GEO} system can overcome the low SNR and meet the decoding condition at the \ac{GEO} satellite. The condition $\alpha < 1$ is possible, but is not practically relevant.} \MG{As illustrated in Fig.~\ref{fig:macframes} case (b),} the slot duration may differ between systems. In this way, we denote with $\nslleo$ and $\nslgeo$ the number of slots in the \ac{MAC} frames for the \ac{LEO} and \ac{GEO} systems, respectively. The two are related according to
%${\frac{\nslleo}{\nslgeo}=\ratefrac}$.
\MG{${{\nslleo}/{\nslgeo}=\ratefrac}$.}
%\begin{equation}
%label{eq:alphansl}
%\frac{\nslleo}{\nslgeo}= \ratefrac
%\end{equation}
%where $\ratefrac$ will be formally defined in Sec.~\ref{sec:PHY} eq.~\eqref{eq:alphadef}. 
The total number of users accessing the wireless medium is $\us=\usleo + \usgeo$, with $\usleo$ and $\usgeo$ being the number of active users served by the \ac{LEO} and \ac{GEO} systems, respectively. We define $\usfrac$ as the ratio between the two. Formally
\begin{equation}
\label{eq:betadef}
\usfrac \triangleq \frac{\usleo}{\usgeo}.
\end{equation}

The channel load of the system is denoted by $\lo$ and corresponds to the average number of \emph{innovative} packets per slot duration.
To correctly compute $\lo$, either the \ac{LEO} or the \ac{GEO} slot duration must be taken as reference. Then, the number of active users needs to be appropriately scaled for the other system, e.g. by taking the \ac{LEO} system perspective:
%
%one must first decide which slot duration is taken as reference, i.e., the \ac{LEO} or \ac{GEO} system slot duration.
%After, one needs to scale accordingly the active users in the two systems. For example, by taking the \ac{LEO} system perspective:
\begin{equation}
\label{eq:loadLeo}
\lo = \frac{\usleo+\ratefrac \usgeo}{\nslleo}.
\end{equation}
In eq.~\eqref{eq:loadLeo} the \ac{GEO} traffic spans $\ratefrac$ \ac{LEO} slots and thus we account for it as $\ratefrac \usgeo$ packets. It can be easily shown that the channel load does not depend on the reference slot duration.
\MG{The receiver attempts the decoding of the replicas only after the reception of an entire \ac{MAC} frame.} Details on the channel model and decoding condition are discussed in Sec.~\ref{sec:PHY}.
%When decoding is successful, the replica content is accessible to the receiver, thus, knowledge about both the slots selected is available.
\MG{A successfully decoded replica provides information about the user slot selections, such that} \ac{SIC} \MG{is able to remove both replicas' contributions from the frame.}
%can be performed and both replicas' contributions can be removed from the frame. In the following,
Ideal interference cancellation \MG{is assumed}, i.e., no residual interference is left after cancellation. Such a procedure is iterated until no further packets can be decoded. In scenario~($\sbb$), the two user populations served by \ac{LEO} and \ac{GEO} satellites coexist over the same bandwidth. In this setting, it is assumed that the receiver on board each satellite is able to perform \ac{SIC} on both traffic types. 
%----------------------------------------------------
\subsection{Channel Model and Decoding Condition}
\label{sec:PHY}

We assume an \ac{AWGN} channel where the attenuation of the transmitted signal is given by free-space path loss, and we consider \ac{TIN}-\ac{SIC} decoding within a slot. Hence, we abstract the physical layer by means of an (asymptotic) outage probability model, where a packet is correctly decoded if and only if
\begin{equation}
\label{eq:deccond}
\rate_\x < \avMutInf_\y,\,\, \text{where } \x,\y\in\{\leo,\geo\}.
\end{equation}
In \eqref{eq:deccond}, the \emph{average instantaneous mutual information}  $\avMutInf$ captures information about the signal-to-noise plus interference ratio experienced by the packet on which we are attempting decoding, hence encapsulating both the interference from other users and the link budget effects. Note that, in \eqref{eq:deccond}, $\x\in\{\leo, \geo\}$ is used to denote whether a packet is served by the \ac{LEO} or \ac{GEO} system. The average instantaneous mutual information must be computed accordingly by setting $\y=\leo$ when we consider
\MG{the receiver on-board the \ac{LEO} satellite and $\y=\geo$, if decoding is done at the \ac{GEO} satellite instead.}
%a receiver's perspective on the \ac{LEO} satellite, and $\y=\geo$ when the perspective is from the \ac{GEO} satellite.
Let us consider a specific replica. We define $\Pw_\y$ and $\Ns_\y$ as the received power and noise power, respectively. Moreover, let us denote by $\nc$  the number of interfering packets over the replica, when decoding a LEO packet, and by  $\nc_j$ (with $j=1,2,\ldots,\alpha$) the number of interfering packets over the $j$th length-$m_\leo$ portion of a packet, when decoding a GEO packet. By assuming perfect power control, i.e., all packets are received with the same power, and with Gaussian codebooks, we have
\begin{subnumcases}{\avMutInf_\y =}
    \log_2\left(1+\dfrac{\Pw_\y}{\Ns_\y+\nc \Pw_\y}\right)  \label{eq:instmutinfo_one}\\ 
    \dfrac{1}{\ratefrac}\displaystyle{\sum_{j=1}^{\ratefrac}}\log_2\left(1+\dfrac{\Pw_\y}{\Ns_\y+\nc_j\Pw_\y}\right) & \label{eq:instmutinfo_two}
\end{subnumcases}
where eq.~\eqref{eq:instmutinfo_one} applies to scenario~($\saa$) and for the \ac{LEO} traffic of scenario~($\sbb$), while eq.~\eqref{eq:instmutinfo_two} applies to the \ac{GEO} traffic of scenario~($\sbb$). Note that for scenario~($\sbb$), the \ac{GEO} traffic coexists with \ac{LEO} traffic. In this way, packet collisions may affect only a portion of the \ac{GEO} packet and we shall take this into account in the average mutual information computation. Additionally, for scenario~($\saa$) it always holds $\x=\y$. For scenario~($\sbb$) instead, the traffic of both systems is received by the \ac{LEO} and by the \ac{GEO}. In the following, the signals and noise powers are computed according to the system parameters summarized in Table~\ref{tab:params2}.
%Note that, the transmit power for all \ac{IoT} devices is identical, irrespective if the desired receiver is the \ac{GEO} or the \ac{LEO} satellite.
\MG{Note that all \ac{IoT} devices transmit with the same power regardless of the targeted receiver.}
The lower \ac{SNR} for the \ac{GEO} receptions is compensated by setting \MG{a lower rate $\rategeo$}.
\MG{This is similar in effect to using repetitions, as is the case of \ac{NB-IoT} \cite{kanj2020tutorial}.}
%Nevertheless, in our model, the prolonged transmission leads to the accumulation of mutual information due to error-control coding.
%The effect of this is similar to the effect of using repetitions in NB-IoT \cite{kanj2020tutorial} in order to accumulate raw signal power at the receiver.

%-----------------------------------------------
\subsection{Performance Metric: Maximum Throughput}
\label{sec:perf_met}

We are comparing the two scenarios in terms of \emph{normalized throughput}~$\GP$, i.e. the average number of successfully received information bits per second and per unit of bandwidth. Formally,
\begin{equation}
    \label{eq:norm_T}
    \GP_\x = \begin{cases}\frac{1}{2} \rate_\x\, \lo_\x\, \psx(\lo_\x, \rate_\x)\quad\text{for scenario~($\saa$)}\\
    \rate_\x\, \lo_\x\, \psx(\lo_\x, \rate_\leo, \rate_\geo, \usfrac)\quad\text{for scenario~($\sbb$)}
    \end{cases}
\end{equation}
with $\x \in \{\leo,\geo\}$ and where $\ps$ is the probability of successful packet reception. In scenario~($\saa$), $\ps$ is a function of the channel load and the selected rate of the \ac{IoT} service. In scenario~($\sbb$) due to the concurrent transmission of the two \ac{IoT} services over the same band, $\ps$ is a function of the load, the rate selected for the \ac{LEO} and \ac{GEO} traffic and $\usfrac$.  For scenario~($\saa$) we account for the fact that each \ac{IoT} service gets a bandwidth $\ban$, effectively being allowed to transmit over only half of the total bandwidth with respect to scenario~($\sbb$). Finally, the maximum throughput for a selected rate is evaluated according to
\begin{equation}
    \label{eq:max_throughput}
    \GPmax_\x = \max_{\lo_\x} \GP_\x.
\end{equation}
Thanks to eq.~\eqref{eq:max_throughput} we can investigate how the maximum throughput changes as we vary the selected rates. 

%% file: analytic_approx.tex
\section{Analytical Approximation of the Maximum Throughput for Scenario~($\saa$)}
\label{sec:an_approx}

In this section, we develop an analytical approximation of the maximum throughput for scenario~($\saa$) by means of \ac{DE}~\cite{Richardson_2001a, Liva2011} analysis. The analysis holds in the limit of large frame sizes. Following~\cite{Liva2011}, we introduce a graphical model of the \ac{SIC} process. In particular, we represent a \ac{CRDSA} \ac{MAC} frame with two sets of nodes: user nodes, associated with active users, and slot nodes, associated with slots. An edge connecting a user node $i$ to a slot node $j$ represents the transmission of one packet replica in the $j$th slot by the $i$th user. The node degree represents the number of edges emanating from one node. For the analysis, it is helpful to define the edge-perspective degree distributions $\ud$ and $\sd$, where $\ud_d$ is the fraction of edges connected to degree-$d$ user nodes, and $\sd_d$ is the fraction of edges connected to degree-$d$ slot nodes. In the limit of large frame sizes, we have that slot node degrees are Poisson-distributed, with
\begin{equation}
\label{eq:rho}
\sd_d = \frac{e^{-2\lo}(2\lo)^{d-1}}{(d-1)!}
\end{equation}
for $d \geq 1$. Moreover, owing to the two repetitions used by \ac{CRDSA}, we have that $\lambda_2 = 1$.
Now we can leverage \ac{DE} so to evaluate the \ac{SIC} performance for asymptotically large frames. We do this by selecting a random user node and developing its neighborhood down to a fixed depth. The resulting graph is tree-like with high probability. We then proceed by evaluating the probability that a user packet is unknown at a given iteration. We refer to the event that decoding of the user packet replica in a slot is unsuccessful as an erasure. We thus track the evolution of the erasure probability over the tree, starting from the leaf nodes, up to the root node (see \cite{Liva2011}).

Let us denote by $q_i$ the probability that, at the $i$th iteration, an edge carries an erasure message from a user node to a slot node. Furthermore, denote by  $p_i$ the probability that, at the $i$th iteration, an edge carries an erasure message from a slot node to a user node. We introduce the recursions
\begin{align}
    q_i &= \fb(p_{i-1}) \label{eq:fb}\\
    p_i &= \fs(q_i;\lo) \label{eq:fs}
\end{align}
which govern the evolution of the erasure probabilities across iterations. Recalling that $\lambda_2 = 1$, we have that \cite{Liva2011}
\begin{equation}
    \fb(p_{i-1}) = p_{i-1}. \label{eq:fb2}
\end{equation}
The derivation of \eqref{eq:fs} requires computing the probability that a user packet replica cannot be decoded over a slot, where $d$ transmissions take place and where each of the transmissions interfering with the packet replica of interest has not been canceled with probability $q_i$. We refer to this probability as $\fs(q_i;d,\lo)$. We have that
\begin{equation}
    \label{eq:pic}
    \fs(q_i;d,\lo) = 1- \sum_{r=0}^{\tau} \binom{d-1}{r} q_i^r (1-q_i)^{d-r-1}
\end{equation}
for $d>\tau$, whereas $\fs(q_i;d,\lo)=0$ if $d \leq\tau$, where $\tau$ is the maximum number of colliding packets that can be successfully decoded. Resorting to the intra-slot \ac{TIN}-\ac{SIC} model of Section~\ref{sec:PHY}, $\tau$ is simply the largest non-negative integer that satisfies
\begin{equation}
    \label{eq:tau}
\rate < \log_2\left(1+\dfrac{\Pw}{\Ns+\tau \Pw}\right).
\end{equation}
The recursion \eqref{eq:fs} can be evaluated by averaging \eqref{eq:pic} over the slot node degree distribution \eqref{eq:rho}. Note that \eqref{eq:pic} resembles the results in \cite{Clazzer2017_IRSA_Fading}.

%--------------------------------------------------------------------------------------------
\subsection{Computation of $\loth$ and Approximation of the Maximum Throughput $\GPmaxAp$}

The \ac{DE} analysis proceeds by setting $q_0=1$, and by iterating \eqref{eq:fb} and \eqref{eq:fs}.
In particular, due to \eqref{eq:fb2}, we can analyze the fixed points of the equation $p_i = \fs(p_{i-1};\lo)$. We denote by $\loth$ the largest value of $\lo$ for which
\begin{equation}
x > \fs(x;\lo) \label{eq:threshold_condition}
\end{equation}
for all $x \in (0,1)$. This is the largest value of the load for which the erasure probabilities become vanishing small for a large number of iterations.  
The approximation of the maximum throughput is obtained by
\begin{equation}
    \label{eq:GPapprox}
    \GPmaxAp = \frac{\rate\, \loth}{2}
\end{equation}
owning to the fact that for scenario~($\saa$) only half of the total bandwidth compared to scenario~($\sbb$) can be utilized. Thanks to the approximation in \eqref{eq:GPapprox}, we are able to evaluate the performance of scenario~($\saa$) without the need for extensive numerical simulations. In the next section, we investigate the tightness of the approximation with respect to finite-length numerical results and we show how it is able to accurately predict the rate that maximizes the peak throughput.

%% file: results.tex
\section{Results}

\begin{figure}
    \centering
    \includegraphics[width=\linewidth]{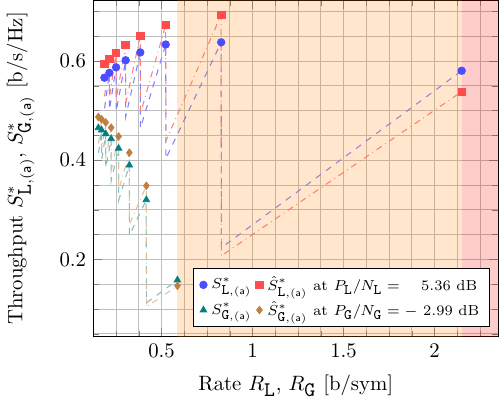}
    \caption{Scenario~($\saa$): maximum throughput of \ac{LEO} traffic received at the \ac{LEO} satellite and of \ac{GEO} traffic at the \ac{GEO} satellite as a function of the rate.  Red and orange areas represent the rate values not permitted due to the link budget constraints for the \ac{LEO} and \ac{GEO} systems, respectively.}
    \label{fig:leogoodput_scenarioA}
    \vspace{-0.3cm}
\end{figure}

\begin{figure}
    \centering
    \includegraphics[width=\linewidth]{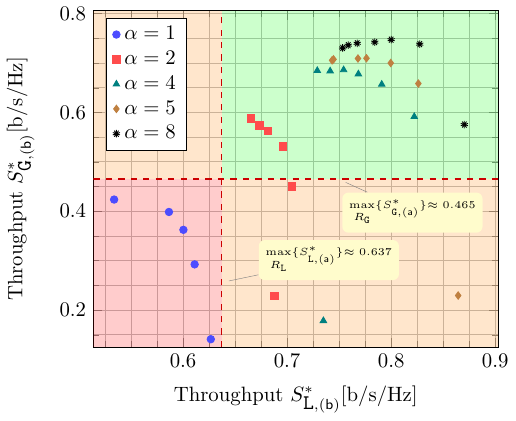}
    \caption{Scenario~($\sbb$): \MG{maximal} throughput pairs for $\usfrac = 1$ when both systems coexist on two channels of total bandwidth $\sysban$.}
    \label{fig:goodput_pairs_beta1}
\end{figure}

Monte Carlo simulations have been conducted for both scenarios by setting the parameters described in the model of Section~\MG{\ref{sec:sys_ov}} according to Table~\ref{tab:params2}. These values are taken from realistic satellite systems, i.e., Inmarsat F2 satellites for \ac{GEO} and OneWeb satellites for \ac{LEO}, whenever available. The \ac{SNR} at the receiver input results in $5.36$~dB \MG{and} $-2.99$~dB for the \ac{LEO} and \ac{GEO} \MG{satellites}, respectively.
\begin{table}
    \centering
    \caption{Simulation Parameters}
    \renewcommand{\arraystretch}{1.2}
    \label{tab:params2}
    \begin{tabular}{ p{0.45\linewidth}  c  c  r }
    \hline\hline \textbf{Description} & \textbf{Symbol} & \textbf{Value} & \textbf{Units} \\
    \hline Transmission power & $\Pw_t$ & 23 & dBmW \\
    \hline User antenna gain & $\AG_t$ & 0 & dBi \\
    \hline \ac{LEO} rx antenna gain & $\AG_{r,\leo}$ & 24.2 & dBi \\
    \hline \ac{GEO} rx antenna gain & $\AG_{r,\geo}$ & 43.6 & dBi \\
    \hline Free space loss for \ac{LEO} system & $\FSL_{\leo}$ & 161.4 & dB \\
    \hline Free space loss for \ac{GEO} system & $\FSL_{\geo}$ & 190.6 & dB \\
    \hline System noise temperature \ac{LEO} & $\SNT_{s,\leo}$ & 26.4 & dBK \\
    \hline System noise temperature \ac{GEO} & $\SNT_{s,\geo}$ & 25.0 & dBK \\
    \hline Bandwidth & $\ban$ & 180 & kHz \\
    \hline Carrier frequency & $\carrfreq$ & 2 & GHz \\
    \hline \ac{CRDSA} frame size for \ac{LEO} & $\nslleo$ & 400 & slots \\
    \hline\hline
    \end{tabular}
\end{table}
Fig.~\ref{fig:leogoodput_scenarioA} shows the maximum throughput as a function of the rate $\rateleo$, $\rategeo$ for each service operating under the assumptions of scenario~($\saa$). The \ac{CRDSA} \ac{MAC} frame has a duration of $\nslleo=\nslgeo=400$~slots and each system uses a dedicated channel of bandwidth $\ban$. Observe that the peak throughput follows a sawtooth profile. This is because we considered a perfect power control scenario and thus the number of colliding packets that can be successfully decoded in the intra-slot \ac{SIC} varies sharply. As a consequence, any rate smaller than $\log_2\left(1+\Pw/(\Ns+\nc\Pw)\right)$ but larger than  $\log_2\left(1+\Pw/(\Ns+(\nc+1)\Pw\right)$ would provide the same performance in terms of successful decoding probability. Hence, the maximum throughput will be directly proportional to the rate for any value within the two aforementioned boundaries. The analytical approximation -- red squares and brown diamonds -- provides a good estimate of the maximum throughput for finite-length \ac{MAC} frames along the entire range of rates (within an $8\%$ error). \MG{Furthermore}, the analytical approximation can predict the rate choice \MG{that maximizes} the throughput. \MG{Note} that the choice differs between the \ac{LEO} and \ac{GEO} systems \MG{due to the dependency} on the \ac{SNR}.
We now investigate the performance of both systems under scenario ($\sbb$), which assumes that each satellite system accesses the medium with \ac{CRDSA} on both originally allocated channels at the same time. The discussion is first constrained to the case in which $\usfrac = 1$, where the common satellite coverage area contains the same number of active users serviced by each system. Fig.~\ref{fig:goodput_pairs_beta1} shows values of maximum throughput grouped according to $\ratefrac$ for each satellite system when simultaneously sharing the entire band. Each value of throughput is achieved for a specific rate pair $\{\rateleo, \rategeo\}$. Furthermore, the maximum achievable throughput for each system under scenario ($\saa$) is depicted with a dashed line and serves as a benchmark for comparison. The two dashed lines divide the result space into four quadrants of interest. The bottom-left quadrant contains throughput values that are unsatisfactory for both satellite operators, as each would incur a penalty with respect to operating separately on its own band. Consequently, for $\ratefrac = 1$, the operators would be better off by not coexisting. The top-right quadrant, however, depicts the opposite. There are multiple values of $\ratefrac$ for which the operators have an incentive to coexist, each surpassing the performance it would achieve by operating on segregated bands. We see that the maximum throughput values are achieved for $\ratefrac = 8$. The remaining quadrants represent a conflict of interest. If a throughput pair is in the bottom-right quadrant, the \ac{LEO} operator would desire coexistence over separation, but exactly the opposite is true for the \ac{GEO} operator. This \MG{holds} for a number of rate pairs corresponding to  $\ratefrac \in \{2, 4, 5\}$. The complementary situation is represented by the top-left quadrant. Here, the \ac{GEO} operator would rather coexist while the \ac{LEO} operator would prefer to use its own band exclusively.
\begin{figure}
    \centering
    \includegraphics[width=\linewidth]{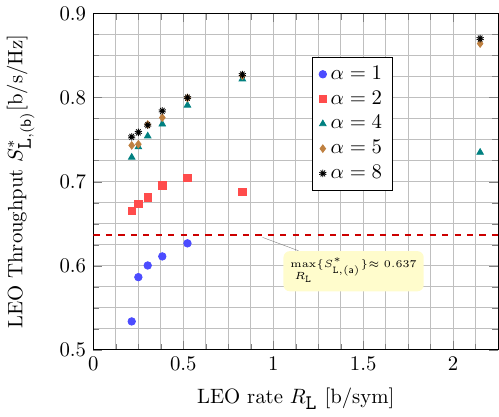}
    \caption{Scenario~($\sbb$): maximum throughput of \ac{LEO} traffic received at the \ac{LEO} satellite for \textbf{$\beta = 1$} as a function of $\rateleo$ and $\ratefrac$}
    \label{fig:leoleo_alp_const}
\end{figure}

\begin{figure}
    \centering
    \includegraphics[width=\linewidth]{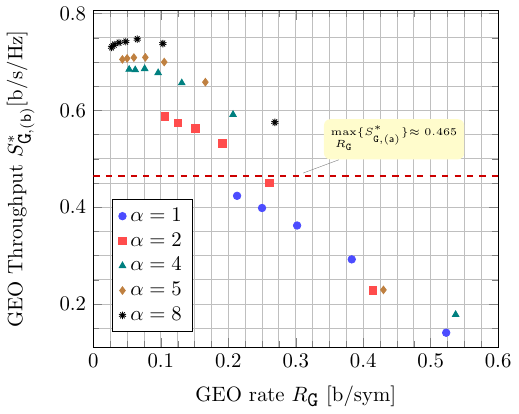}
    \caption{Scenario~($\sbb$): Maximum throughput of \ac{GEO} traffic received at the \ac{GEO} satellite for \textbf{$\beta = 1$} as a function of $\rategeo$ and $\ratefrac$.}
    \label{fig:geogeo_alp_const}
\end{figure}

In Fig.~\ref{fig:leoleo_alp_const} we inspect the achievable throughput for \ac{LEO} users for $\usfrac=1$. For each $\ratefrac$, there exists a rate choice that maximizes the throughput. Furthermore, despite coexisting with the \ac{GEO} system, we see that the throughput for the \ac{LEO} system peaks at $0.870$~[b/s/Hz]. This represents an increase of $36.5\%$ in peak throughput over scenario ($\saa$). Similarly, the \ac{GEO} operator's performance is captured by Fig.~\ref{fig:geogeo_alp_const}, which shows that the throughput values are trending roughly inversely proportional to the rate values available for the \ac{GEO} users. A peak throughput of $0.747$~[b/s/Hz] is also achieved for $\ratefrac = 8$, which marks a gain of $60.6\%$ over the peak throughput value of scenario ($\saa$). While each operator can agree to choose a rate that results in $\ratefrac=8$, it does not achieve maximum throughput for the same rate pair as the competitor. A mechanism needs to be devised to let the operators agree on a suitable rate pair.

\begin{figure}
    \centering
    \includegraphics[width=\linewidth]{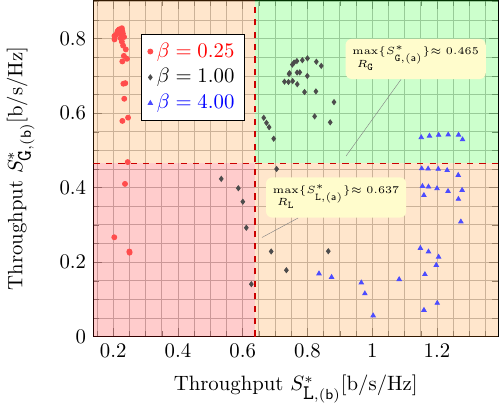}
    \caption{Scenario~($\sbb$): maximum throughput pairs for $\usfrac=1/4,~\usfrac=1$ and $\usfrac=4$ when both systems coexist on two channels of joint bandwidth $\sysban$.}
    \label{fig:allgoodputpairs}
\end{figure}

We now expand the discussion to include results for a case where more active users are serviced by the \ac{LEO} operator than the \ac{GEO} one, i.e., for $\usfrac = 4$; and results for the mirrored case, namely for $\usfrac = 0.25$. Fig.~\ref{fig:allgoodputpairs} shows three sets of throughput pairs corresponding to different $\usfrac$. Red circular markers correspond to $\usfrac=0.25$, black diamond markers to $\usfrac=1$, and blue triangular markers to $\usfrac=4$.
%Each throughput pair is achieved for a different pair of \ac{LEO} and \ac{GEO} rates}.
This figure is also divided into four quadrants with the same meaning as per Fig.~\ref{fig:goodput_pairs_beta1}. Should the number of users served by the \ac{GEO} operator be four times the number serviced by the \ac{LEO} operator ($\usfrac = 0.25$), then there are no throughput pairs available that would incentivize the \ac{LEO} operator to choose coexistence, as all throughput values are lower than the peak throughput of scenario~($\saa$). The \ac{GEO} operator, on the other hand, has points available that express a desire for coexistence (top left quadrant). Finally, should $\usfrac=4$, a case where there are four times as many \ac{LEO} users than \ac{GEO} users, then the \ac{LEO} operator would always prefer to coexist with the \ac{GEO} (top-right and bottom-right quadrants). The mutual desire to coexist, however, is given when $\ratefrac=8$. This latter case is represented by throughput pairs marked with blue triangles, which are exclusively located in the top-right quadrant.
